# Avoiding Aliasing in Allan Variance: an Application to Fiber Link Data Analysis

Claudio E. Calosso, Cecilia Clivati, Salvatore Micalizio
Physics Metrology Division
Istituto Nazionale di Ricerca Metrologica, INRIM
Torino, Italy
e-mail: c.calosso@inrim.it

*Abstract* — Optical fiber links are known as the most performing tools to transfer ultrastable frequency reference signals. However, these signals are affected by phase noise up to bandwidths of several kilohertz and a careful data processing strategy is required to properly estimate the uncertainty. This aspect is often overlooked and a number of approaches have been proposed to implicitly deal with it. Here, we face this issue in terms of aliasing and show how typical tools of signal analysis can be adapted to the evaluation of optical fiber links performance. In this way, it is possible to use the Allan variance as estimator of stability and there is no need to introduce other estimators. The general rules we derive can be extended to all optical links. As an example, we apply this method to the experimental data we obtained on a 1284 km coherent optical link for frequency dissemination, which we realized in Italy.

## I. INTRODUCTION

In recent years, coherent optical fiber links have become a well established tool for frequency dissemination [1-5] and a growing number of fiber-based atomic clock comparisons are going to be performed in next years [6-8]. This is motivated by the increased accuracy and stability of this technique as compared to the state-of-the-art of satellite links: optical links achieve a resolution at the level of $1\times10^{-19}$ after one day of averaging time and are likely to replace satellite techniques at least on continental scales. This opens new possibilities not only in metrology, but also in fundamental physics [9], high-precision spectroscopy [10], geodesy [11] and radio-astronomy [12].

Coherent frequency transfer via optical fiber is based on the delivery of an ultrastable laser at telecom wavelength along a standard fiber for telecommunications, where the length variations are actively canceled through the Doppler-noise-cancellation scheme [13] or rejected by means of the 2-way technique [14, 15]. With these schemes, the intrinsic noise of thousands-kilometers long fibers due to vibrations and temperature changes can be reduced up to a bandwidth of tens of hertz. The residual phase noise of the delivered laser actually depends on the intrinsic noise of the fiber itself and on the fiber length [16, 17]. The several experimental realizations so far implemented all have some common features: the compensation bandwidth $B_L$ is typically few tens of hertz, determined by the link length $L$ through the relation $B_L = c_n/(4L)$ ($c_n$ being the speed of light in the fiber). In addition, assuming that the phase noise can be described by the power spectral density law $S_\varphi(f)=\sum_\alpha b_\alpha f^\alpha$, ($-4 \leq \alpha \leq +1$), the link phase noise is most often of the type $S_\varphi(f) \propto f^0$ (white phase noise, WPN) or $S_\varphi(f) \propto f^1$ (blue phase noise, BPN) [2, 3]. At very low Fourier frequencies ($f < 0.01$ Hz), the uncompensated fiber noise is negligible as compared to some long-term effects on the interferometer and to the intrinsic noise of the clocks to be compared. In addition, a strong bump has been reported in many realizations [1-4] between 10 Hz and 30 Hz. This is due to uncompensated acoustic noise and vibrations on the optical fiber and to the servo.

Of course, one is interested in rejecting such high-frequency noise, since, in general, it does not contain useful information. Although these noise components can be effectively discriminated in a frequency-domain measurement, it is not straightforward to identify and reject them in a time-domain measurement. The resulting effect is well known, for instance, when evaluating the performance of optical links with frequency counters. Specifically, if a proper procedure is not adopted, the traditional estimator for statistical uncertainty, the Allan variance (AVAR) [18], is saturated by the link high-frequency noise. However, it would be desirable to keep the use of AVAR also in this case, since it has been recognized by the metrological community as the estimator for instability in presence of colored noise processes.

In this work, we show that this problem can be stated in terms of aliasing. Aliasing is often described in the frequency domain through the spectral folding of fast Fourier components in the low frequency region of the spectrum and causes a degradation of the sampled signal [19, 20]. As a result, any further analysis on it will be affected, either in the frequency or in the time domain, including the AVAR estimation. This problem can be avoided quite straightforwardly if sampling time and measurement bandwidth are properly chosen. Under this assumption, we show how the AVAR can be used as an estimator for instability, allowing a direct and unambiguous comparison of the results, in continuity with the existing literature.

We recall that the AVAR is indeed expressed as a function of the measurement bandwidth $f_h$. However, this degree of freedom is not commonly exploited as a means to reject high-frequency noise or aliasing; on the contrary, it has become a common practice to use the Modified Allan Variance (MVAR) as an estimator for frequency stability in optical links [2, 4, 8]. The MVAR has the advantage of mitigating the effect of the link high-frequency noise already at short averaging times, but, as a drawback, the estimation of the long-term instability is affected as well. In the case of white frequency noise (WFN), which is the typical case when atomic clocks are compared by means of an optical link, the MVAR no longer corresponds to the classical variance [21-23]. We stress that this correspondence is the underlying reason behind the choice of the AVAR as a stability estimator in the metrological community [21]. Thus, the choice of the MVAR has to be motivated, and recent works are addressing this issue [24].

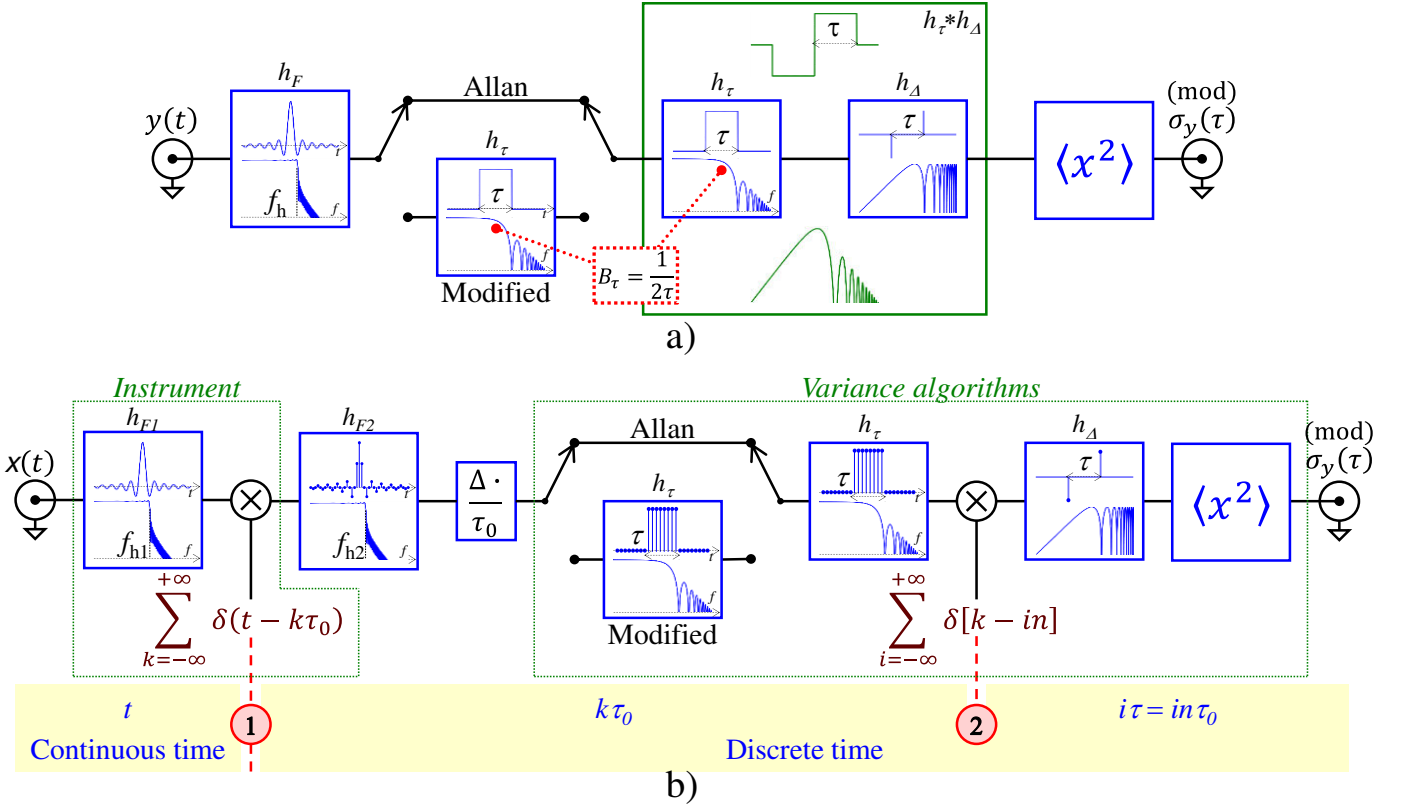

Fig. 1: Functional block diagram of the AVAR and of the MVAR. a) highlights the role of the measurement bandwidth $f_h$ represented by the first block $h_F$ (both impulse response and transfer function are shown); the switches allow selection between MVAR and AVAR, by enabling/disabling the additional moving average filter $h_\tau$. Both $h_F$ and $h_\tau$ are low-pass filters that reject fast noise, but the MVAR additional $h_\tau$ is less selective and changes its bandwidth with $\tau$. The green block represents the well-known AVAR equivalent filter, i.e. the average $h_\tau$ over a time $\tau$ and the difference $h_\Delta$ between adjacent samples divided by $\sqrt{2}$; the last block evaluates the mean power. b) shows the two stages where aliasing may occur: 1) the measurement instrument that samples the phase time $x(t)$ each $\tau_0$ and 2) the AVAR and MVAR algorithms that decimate the average fractional frequency by $n$, reducing the data rate from $1/\tau_0$ to $1/\tau$. Because of sampling the base time is discrete. The additional low-pass filter $h_{F2}$ can be used to close further the measurement bandwidth. $h_F$ groups $h_{F1}$ and $h_{F2}$. The $\Delta \cdot/\tau_0$ block differentiates the phase time to obtain the fractional frequency.

In the following sections, we will describe our approach; in particular, we will separate the measurement process from the AVAR computation and propose a theoretical and experimental procedure to avoid aliasing when using the most common acquisition instruments: although phasemeters would be the best choice, thanks to their selective anti-aliasing filter [25], also commercial phase and frequency counting devices can be, to some extent, adapted to this task [26]. We then propose an experimental case where this method can be applied, showing the results we obtained on the 1284 km optical fiber link for frequency dissemination developed by the National Metrology Institute in Italy (INRIM) [3].

## II. THEORY

We suppose to work with an optical frequency link and we want to characterize its performance; this is routinely made by comparing two links with the same starting and ending points [4] or by looping a single link, so that the remote end coincides with the local end [1-3]. Similarly, we can suppose to compare/disseminate remote clocks.

In both cases, the output of the measurement is a stream of dead-time free phase or frequency data that are processed by statistical tools. Every measurement system has a finite bandwidth that can be modeled by a low-pass filter with an equivalent bandwidth $f_h$. Ideally, this filter completely rejects the noise above $f_h$ and can be either hardware or software realized. The data are then processed by a statistical estimator.

Specifically, we focus on the two typical signal analysis methods commonly adopted in frequency metrology, the AVAR and the MVAR.

The AVAR is a well-known estimator and is universally accepted in frequency metrology. It is described by the following formula:

$$\sigma_y^2(\tau) = \frac{1}{2} \left\langle \left(\bar{y}(t) - \bar{y}(t-\tau)\right)^2 \right\rangle$$
$$= \left\langle \left(y(t) * h_F(t) * h_\tau(t) * h_\Delta(t)\right)^2 \right\rangle$$
$$= \int_0^\infty S_y(f) |H_F(f) H_\tau(f) H_\Delta(f)|^2 \, df \quad (1)$$

where $\bar{y}(t)$ represents the average fractional frequency over the time interval $[t-\tau, t]$ and $\tau$ is the averaging time; the brackets $\langle ... \rangle$ denote an infinite time average; $h_F$, $h_\tau$, $h_\Delta$ are the impulse responses of the low-pass filter that sets the measurement bandwidth, of the average over $\tau$ and of the difference divided by $\sqrt{2}$ respectively. The star stands for the convolution product. In the last line of (1), we wrote its expression in the frequency domain, where $S_y(f) = \left(\frac{f}{\nu_0}\right)^2 S_\varphi(f)$ ($\nu_0$ is the carrier, $\varphi$ is the phase of the signal) and $H_F$, $H_\tau$, $H_\Delta$ are the Fourier transforms of $h_F$, $h_\tau$, $h_\Delta$ respectively.

Historically, the MVAR was introduced to deal with fast noise processes, such as WPN. It is defined by the formula:

$$\text{Mod } \sigma_y^2(\tau) = \frac{1}{2} \left\langle \left(\frac{1}{n}\sum_{i=1}^n \bar{y}(t-i\tau_0) - \frac{1}{n}\sum_{i=1}^n \bar{y}(t-\tau-i\tau_0)\right)^2 \right\rangle$$
$$\cong \left\langle \left(y(t) * h_f(t) * h_\tau(t) * h_\tau(t) * h_\Delta(t)\right)^2 \right\rangle$$
$$= \int_0^\infty S_y(f) |H_F(f) H_\tau^2(f) H_\Delta(f)|^2 \, df \quad (2)$$

Similarly to (1), we also expressed it in the frequency domain. The approximation in (2) holds for $\tau \gg \tau_0$, where $\tau_0 = \tau/n$ represents the shortest averaging time, i.e. the original gate time of the measurement instrument.

| Noise type | $S_\varphi(f)$ | $\sigma_y^2(\tau)$ | Mod $\sigma_y^2(\tau)$ | $\frac{\text{MVAR}}{\text{AVAR}}$ | $\frac{\text{MVAR}}{\text{AVAR}}$, dB |
|---|---|---|---|---|---|
| Large bump $B_b > 1/\tau$ | $b_b$ $\|f - f_b\| < \frac{B_b}{2}$ $P_b = b_b B_b$ $f_h > f_b + B_b/2$ | $\frac{3}{4\pi^2} \frac{P_b}{\nu_0^2} \tau^{-2}$ | $\frac{5}{8\pi^4(f_b^2 - B_b^2/4)} \frac{P_b}{\nu_0^2} \tau^{-4}$ | $1.1 \times 10^{-5\S}$ | $-49.5^\S$ |
| Narrow bump $B_b < 1/\tau$ | | $\frac{2}{\pi^2} \sin^4(\pi f_b \tau) \frac{P_b}{\nu_0^2} \tau^{-2}$ | $\frac{2}{\pi^4 f_b^2} \sin^6(\pi f_b \tau) \frac{P_b}{\nu_0^2} \tau^{-4}$ | - | - |
| Blue PN | $b_1 f^1$ | $\frac{3 f_h^2}{8\pi^2} \frac{b_1}{\nu_0^2} \tau^{-2}$ | $\frac{9.643 + 10 \ln(\pi f_h \tau)}{16\pi^4} \frac{b_1}{\nu_0^2} \tau^{-4}$ | $2.6 \times 10^{-5*}$ | $-45.9^*$ |
| White PN | $b_0$ | $\frac{3 f_h}{4\pi^2} \frac{b_0}{\nu_0^2} \tau^{-2}$ | $\frac{3}{8\pi^2} \frac{b_0}{\nu_0^2} \tau^{-3}$ | $0.005^*$ | $-23.0^*$ |
| Flicker PN | $b_{-1} f^{-1}$ | $\frac{1.038 + 3\ln(2\pi f_h \tau)}{4\pi^2} \frac{b_{-1}}{\nu_0^2} \tau^{-2}$ | $\frac{3 \ln(256/27)}{8\pi^2} \frac{b_{-1}}{\nu_0^2} \tau^{-2}$ | $0.166^*$ | $-7.8^*$ |
| White FN | $b_{-2} f^{-2}$ | $\frac{1}{2} \frac{b_{-2}}{\nu_0^2} \tau^{-1}$ | $\frac{1}{4} \frac{b_{-2}}{\nu_0^2} \tau^{-1}$ | $0.5$ | $-3.0$ |
| Flicker FN | $b_{-3} f^{-3}$ | $2 \ln(2) \frac{b_{-3}}{\nu_0^2}$ | $\frac{27}{20} \ln(2) \frac{b_{-3}}{\nu_0^2}$ | $0.675$ | $-1.7$ |
| Random Walk FN | $b_{-4} f^{-4}$ | $\frac{2\pi^2}{3} \frac{b_{-4}}{\nu_0^2} \tau$ | $\frac{11\pi^2}{20} \frac{b_{-4}}{\nu_0^2} \tau$ | $0.825$ | $-0.8$ |
| Linear frequency drift $\dot{y}$ | - | $\frac{1}{2} (\dot{y})^2 \tau^2$ | $\frac{1}{2} (\dot{y})^2 \tau^2$ | $1$ | $0$ |

Table 1: AVAR and MVAR for several noise processes; these formulae hold for $f_h \tau \gg 1$ and $n \gg 1$. $\nu_0$ is the frequency of the optical carrier; $f_h$ is the measurement bandwidth; the bump parameters: $P_b$, $f_b$, $B_b$ and $b_b$ are defined in Sec. II.D. The ratio MVAR/AVAR has been calculated for $f_h \tau = 100$ (*) and for $f_b \tau = 100$ and $f_b = B_b$ (§);

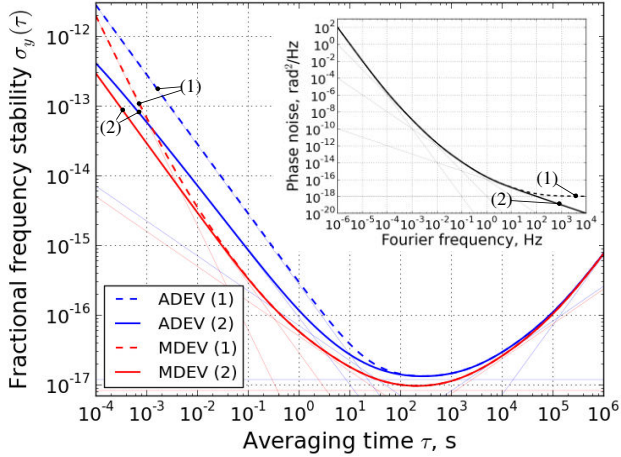

Fig. 2: ADEV (blue) and MDEV (red) provide different estimations, especially for fast noise processes. Here it is shown for the phase noise in the inset, with (dashed line) and without (continuous line) WPN.

The block diagram of the two estimators is sketched in Fig. 1.a. It can be seen that the MVAR is obtained from the AVAR with the introduction of the additional moving-average filter $h_\tau$, the same that is already used in the AVAR computation.

Table 1 reports the analytical derivation of AVAR and MVAR for the various noise types. We extended the analytical formulae commonly found in the literature to the cases of BPN ($\alpha = +1$) [2, 3] and bumps which have often been reported at acoustic frequencies. The latter case will be explicitly addressed in Sec. II.D. In addition, we calculated the ratio MVAR/AVAR.

### A. AVAR vs MVAR

If we compare the two estimators in the frequency domain, it is clear to see that the MVAR transfer function applies twice the first-order low-pass filter $h_\tau$ with the goal to reject high-frequency noise. However, unlike $h_F$ whose bandwidth $f_h$ is fixed, the additional moving average of the MVAR changes its equivalent bandwidth $B_\tau$ with the measurement time $\tau$, according to the relation $B_\tau = 1/(2\tau)$. This is the reason for the strong $\tau^{-4}$ dependence on the MVAR. Most importantly, this additional moving-average filter causes a discrepancy between the results obtained with AVAR and MVAR. In fact, although $h_\tau$ is introduced to reject high-frequency noise, it acts at all timescales. In particular, it also acts at long measurement times where, in most cases, it is not necessary, for processes with $\alpha \leq -2$. Tab. 1 and Fig. 2 show that the faster is the noise process, the higher is the discrepancy between AVAR and MVAR.

Tab. 1 and Fig. 2 summarize this behavior for the various noise types in the case $f_h \tau \gg 1$. It can be seen that MVAR is unaffected in presence of a linear frequency drift, while the discrepancy increases with $\alpha$, i.e. with fast noise processes. In presence of WFN, it is as large as $-50\%$ ($-3$ dB) for all $\tau$. We stress that WFN is the typical behavior of atomic clocks and in this case the discrepancy can indeed be considered as a bias, since the AVAR is universally accepted as estimator for statistical uncertainty. Such bias is an issue because the uncertainty can no longer be related to the standard deviation as expected.

It then appears to be a natural choice to use the AVAR also for the characterization of optical links, since they are meant for the remote comparison and dissemination of atomic clocks.

Another consequence of the MVAR additional filter is that, in order to be applied over a long measurement time, it must have a short impulse response, which implies a poor selectivity. The moving average minimizes the impulse response at the expense of the selectivity, i.e. the ability to reject high-frequency noise. Although the results are only slightly affected in the case of BPN, this is a real limitation in case of bumps.

As an alternative to MVAR, we propose to use the AVAR associated to the low-pass filter $h_F$ already introduced. Such filter is capable of rejecting completely the high frequency noise if its fixed bandwidth $f_h$ and its attenuation are properly set, thus avoiding aliasing induced by the fast noise processes on the AVAR algorithm.

On the other hand, the filter does not bias the long-term behavior of the estimator. This allows an unambiguous comparison of the results among different experiments and with the literature, where the AVAR has been traditionally adopted.

### B. Instrumental and variance aliasing

To see how aliasing affects the AVAR estimation [27], we need to refine the model by considering the entire chain from the measurement of the signal to the computation of the variance. The former is performed in the continuous time domain, while the latter is done in discrete time domain. With reference to Fig. 1.b, we see two points where aliasing occurs: the first one is represented by the instrument that, each $\tau_0$, samples the continuous signal and converts it into a sequence of data. Due to aliasing the high frequency noise of the signal is converted at frequencies lower than $1/(2\tau_0)$ and causes a degradation of the PSD of the sampled signal with respect to the original one in continuous time. Hence, AVAR and MVAR estimations degrade according to the formulae of Tab. 1. For instance, it is well known that in case of WPN, $b_0$ is degraded by a factor $1/(2\tau_0 B)$ ($B$ is the bandwidth of the signal) [28]. The degradation is even larger in the case of fiber links, which are affected by faster noise processes.

The instrumental aliasing can be avoided by simply fulfilling the Nyquist theorem, i.e. $1/\tau_0 > 2B$. In case the instrument sampling speed is limited, a possible way to fulfill the Nyquist theorem is to limit its measurement bandwidth $f_{h1}$, so that $f_{h1} < 1/(2\tau_0)$. This operation in some cases may be not trivial as we will see in Sec. II.A.

The second point where aliasing occurs is less evident and is hidden in the AVAR algorithm itself. To calculate the difference of two adjacent average frequencies, the algorithm reduces the data rate by a factor of $n$, from $1/\tau_0$ to $1/\tau$. In practice, it samples the average frequency $\bar{y}$ each $\tau$. However, as shown by $h_\tau$ in the dashed box of Fig. 1.b, the Allan variance algorithm considers averaging over $\tau$, that acts as anti-aliasing filter with an equivalent bandwidth $1/(2\tau)$ as required by the Nyquist theorem. Again, simple averaging has a 20 dB/dec roll-off, which means poor selectivity. For this reason, such filtering is effective only for $\alpha \leq -2$ and, for faster noise processes, the AVAR is affected by aliasing. This is in agreement with Tab. 1, where it can be seen that the AVAR depends on $f_h$ for $\alpha > -2$; with flicker phase noise (FPN) the dependence on $f_h$ is weak and the AVAR can still be used as an estimator, provided that $f_h$ is indicated; for WPN, BPN or bumps, the contribution of aliasing is strong. In such cases, the AVAR is not representative of the true noise processes at the considered timescale but rather it gives information about what happens at frequency of the order of $f_h$.

As already discussed, in the MVAR algorithm the filter $h_\tau$ is applied twice: this results in a second order anti-aliasing filter, represented by $h_\tau*h_\tau$. It has an equivalent bandwidth of $1/(3\tau)$, that fulfills the Nyquist theorem, and a roll off of 40 dB/dec. Thanks to such higher selectivity, the MVAR is free from aliasing for noise process up to WPN. For BPN, the aliasing is weak and the MVAR can still be used. Nevertheless, for faster noise processes, in particular in presence of bumps, also MVAR suffers for aliasing, although less severely than AVAR.

*C. Optimal measurement bandwidth*

In a real measurement, the spectrum is the sum of many contributions. To have an estimation free from aliasing we have, of course, to filter out the noise processes responsible for that, i.e. those with $\alpha > -2$ if the AVAR is to be used, and those with $\alpha > 0$ if the MVAR is to be used. From a practical point of view, FPN for AVAR and BPN for MVAR might not be filtered out, as they weakly contribute to aliasing.

It is useful to determine analytically on which criteria $f_h$ should be chosen. Before proceeding with a formal statement, we point out that the AVAR, as well as the MVAR, has no meaning for averaging times $\tau < 1/(2f_h)$ [27] because of the filtering procedure. As an example, Fig. 3 shows the role of $f_h$ on AVAR in the case of WFN. The AVAR is underestimated for $\tau f_h \ll 1$ and starts to be representative of the noise process for $\tau = 1/(2f_h)$, where the underestimation is 36% (1.9 dB).

Thus, if $f_h$ is kept too low, the fast noises are completely rejected, but also the aliasing-free components of the spectrum are filtered out and the stability is underestimated. On the other hand, if $f_h$ is kept too high, the aliasing induced by the fast noise processes in the AVAR (MVAR) computation hides the "interesting" information contained in the low-frequency spectrum. As a result, the stability is overestimated. On this basis, we can choose as a criterion to determine the optimal bandwidth $f_{h\mathrm{opt}}$ the frequency where there is no estimation error. This can be achieved by compensating the underestimation of the aliasing-free components with the overestimation induced by the aliasing-affected components. To clarify this point, let us consider the transmission of a clock signal over a fiber link. Here, the "interesting" information is related to the clock signal; its phase noise spectrum is referred as $S_\varphi^{clock}(f)$. The alias-affected component is mainly caused by the link noise, here indicated as $S_\varphi^{link}(f)$.

The previous statement can be mathematically expressed as:

$$\int_{f_h}^\infty S_\varphi^{clock}(f) \frac{f^2}{\nu_0^2}|H_A(f)|^2 \mathrm{d}f = \int_0^{f_h} S_\varphi^{link} \frac{f^2}{\nu_0^2}(f)|H_A(f)|^2 \mathrm{d}f \quad (3)$$

where $|H_A(f)|^2 = |H_\tau(f)H_\Delta(f)|^2 = 2\frac{\sin^4(\pi f \tau)}{(\pi f \tau)^2}$.

From (3) it is interesting to see that in the case of WFN for $S_\varphi^{clock}$ and WPN for $S_\varphi^{link}$, the optimal measurement bandwidth $f_{h\mathrm{opt}}$ coincides with the crossing point $f_\times$ of these spectra. In general, for $S_\varphi^{link} = b_\alpha f^\alpha, \alpha > 0$, $f_{h\mathrm{opt}}$ is slightly higher, being

$$f_{h\mathrm{opt}} = \sqrt[(\alpha+2)]{\alpha+1}\, f_\times < 1.33\, f_\times \quad (4)$$

For simplicity, $f_\times$ can still be considered as optimum within a good approximation, allowing for an efficient visual diagnosis.

The same procedure can be followed, for instance, to characterize the optical link itself with the AVAR. Here, the measurement bandwidth should be chosen as the frequency where the spectrum changes slope from $\alpha = -1$ to $\alpha = 0$.

The theory developed here is general and can be applied in other fields, i.e. to analyze TWSTFT data [29]; nevertheless, the presented concepts are evident in the case of fiber links, thanks to the peculiar nature of the signals. First of all, in optical links the phase noise is much higher than the instrument noise, due to the leverage between optical and RF frequencies. Therefore, the instrument can be regarded as noiseless.

In addition, the signal is band-limited and the Nyquist theorem can be satisfied by most common measurement instruments without any filtering. As will be discussed in Sec. III, this is the underlying reason why counters can be successfully used. In order to treat properly the technical aspects, we need to model the fiber link noise, in particular the characteristic bump, and define the filter that rejects its contribution to AVAR.

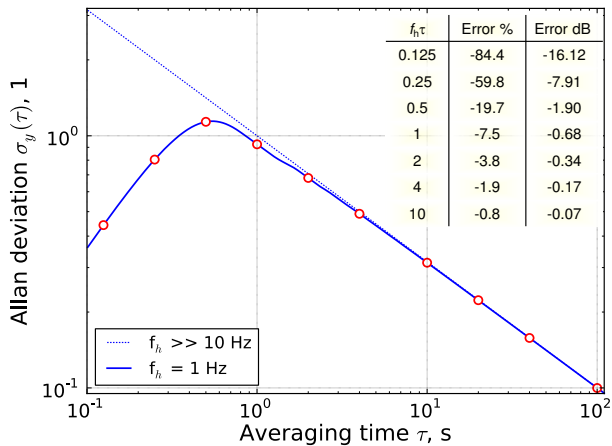

Fig. 3: Filtering effect on the ADEV, the square root of AVAR: the ADEV can be used for $\tau > 1/(2f_h)$. Compared to an almost unfiltered case ($f_h \gg 10$ Hz), the underestimation decreases quickly with $\tau$. The inset table shows the underestimation for various values of $f_h\tau$.

## D. Equivalent bump

The bump that characterizes the unsuppressed noise in fiber links is specific for each realization. It depends on the length of the link, on its noise and on the servo compensating it. The shape is variable and its modeling is not trivial. Nevertheless, we aim at developing a model able to predict the bump contribution to AVAR and MVAR. This can be done successfully by defining the equivalent bump characterized by a rectangular shape with the same power $P_b$, the same average frequency $f_b$, the same bandwidth $B_b$ and the same amplitude $b_b$ of the original bump, where

$$P_b \stackrel{\text{def}}{=} \int_{f_l}^{f_u} S_\varphi(f)\, df$$
$$f_b \stackrel{\text{def}}{=} \frac{1}{P_b}\int_{f_l}^{f_u} f S_\varphi(f)\, df$$
$$B_b^2 \stackrel{\text{def}}{=} \frac{12}{P_b}\int_{f_l}^{f_u}(f-f_b)^2 f S_\varphi(f)\, df$$
$$b_b \stackrel{\text{def}}{=} P_b/B_b$$
(5)

and $f_l$ and $f_u$ are the lower and upper boundaries of the integration domain containing the bump. Depending on the bump bandwidth compared to the averaging time, we can distinguish between large bump ($B_b > 1/\tau$) and narrow bump ($B_b < 1/\tau$). Most of the bumps encountered in optical links at acoustic frequencies fall within the large-bump approximation. Narrow bumps can be used to describe the peak induced by a very high gain servo.

The contribution of the equivalent bump to the AVAR and to the MVAR can be written as

$$\sigma_y^2(\tau) = \int_{B_b} b_b \frac{f^2}{\nu_0^2} |H_A(f)|^2 df$$
(6)

$$\text{Mod } \sigma_y^2(\tau) = \int_{B_b} b_b \frac{f^2}{\nu_0^2} |H_M(f)|^2 df$$
(7)

where $|H_M(f)|^2 = \left|H_\tau^2(f) H_\Delta(f)\right|^2 = 2\frac{\sin^6(\pi f \tau)}{(\pi f \tau)^4}$ and the integration is performed over the interval $[f_b - B_b/2, f_b + B_b/2]$. The results are reported in Tab. 1 for the two cases of large and narrow bumps.

As the other fast noises, the bump is a source of aliasing and must be filtered out by the anti-aliasing filter $h_F$.

## E. Anti-aliasing filter

The optimal cut-off frequency of the anti-aliasing filter has been defined in Sec. II.C. There we implicitly assumed an ideal low-pass filter, with infinite attenuation for $f > f_h$. In real situations, a critical attenuation should be defined, which reduces the contribution of the noise above $f_h$ (that is dominated by the bump) at a negligible level in the variance computation. For the AVAR, this means

$$\int_{B_b} b_b \frac{f^2}{\nu_0^2} |H_A(f)|^2 |H_F(f)|^2 df \ll \int_0^{f_h} S_\varphi(f) \frac{f^2}{\nu_0^2} |H_A(f)|^2 df$$
(8)

For $\alpha \geq 0$ and, again, for $f_h \tau \gg 1$ this can be simplified in:

$$P_b \frac{1}{A_F} \ll \int_0^{f_h} S_\varphi(f) df$$
(9)

where $A_F$ is the average attenuation of the anti-aliasing filter $h_F$ defined as

$$\frac{1}{A_F} \stackrel{\text{def}}{=} \frac{1}{B_b}\int_{B_b} |H_F(f)|^2 df$$
(10)

Equation (4) and (9) directly relate the anti-aliasing filter $h_F$ parameters to the link phase noise spectrum, allowing for an immediate and efficient design. We point out that the reduction of $f_h$ relaxes the specification of the filter, because the transition bandwidth is wider and it is easier for the filter to reach the attenuation $A_F$ specified in (9).

## III. COUNTING IN PRACTICE

It may be useful at this point to give a simple procedure for data acquisition in an experimental session devoted to assess the optical link performance or to compare two remote clocks. In both cases, the variable of interest is phase, as it is proportional to the optical length and it is directly subject to discrete cycle slips. It is then advisable to measure phase rather than frequency; the latter can always be derived from phase data unambiguously.

### A. Measurement instrument

The first consideration is on which instrument to use. Acquisition devices can be grouped into two big families: phasemeters and frequency counters; the difference is mainly related to $h_{F1}$. In phasemeters, the measurement bandwidth $f_{h1}$ is well defined and fulfills the Nyquist theorem, i.e. $f_{h1} < 1/(2\tau_0)$. In addition, $h_{F1}$ is very selective and allows, if properly set, to filter out the bump completely. As a result, the instrument aliasing is automatically avoided in any case, also when $1/\tau_0 < 2B$ (in practice, $\tau_0$ can be set of the order of 1 s).

On the other hand, electronic counters are more common and cheaper. They can be used for data acquisition if dead-time free or if used in the time-interval mode, which is intrinsically dead-time free. However, we note that a good knowledge of the counter operation is not trivial, as demonstrated by the broad literature [30, 31] and care must be taken to avoid incurring in mistakes. For what concerns this work, the main feature of electronic counters is the large bandwidth $f_{h1}$ of the input stage, typically hundreds of megahertz, as compared to the sampling rate, i.e. $f_{h1} \gg 1/(2\tau_0)$. Therefore, the Nyquist theorem is not fulfilled in the general case and the only way to avoid instrumental aliasing is to sample faster than twice the noise bandwidth $B$, i.e. $1/\tau_0 > 2B$. It is important to note that in this case, contrary to the general opinion, it is totally irrelevant which kind of counter is used.

The sampling rate depends on the counter model: for instance, Stanford Research SR620 or any counter used in the so-called Π-mode has a sampling frequency $f_s = 1/\tau_0$; Keysight (formerly Agilent) 53131/2 has $f_s$ = 200 kHz; Kramer & Klische FXE80 has $f_s$ = 1 kHz (or less); Guidetech GT668 has $f_s$ = 4 MHz. Most of the counters with high sampling rate implement some sort of filtering before decimating to $1/\tau_0$, in order to reduce the contribution of aliasing. This is most often done by applying a simple moving average filter (the so-called Λ-mode), as it requires minimal computational effort. Emerging techniques are based on weighted averaging (e.g. the so-called Ω-mode, based on the least-squares interpolation) [31]. In any case, these filters are not selective enough to reject the bump completely and aliasing is only reduced, not avoided. This is particularly evident in the case $f_h \approx f_b$. The implementation of more selective low-pass filters with programmable bandwidth would allow to reduce the sample rate of the counter below $2B$ and the amount of data to be acquired. At the moment, the only way to avoid instrumental aliasing with counters in case of fiber links is to set the gate time $\tau_0$ below $1/(2B)$, in practice $\tau_0 <$ 10 ms for thousands-kilometers links.

*B. Filtering*

In Sec. III.A we dealt with the rejection of the instrumental aliasing. The next step is the design of the $h_F$ filter that rejects the variance aliasing. This means to choose properly the optimal bandwidth $f_{hopt}$ and the attenuation $A_{F\min}$. All the required information are contained in the phase noise power spectrum. As seen in Sec. II.C, $f_{hopt}$ can be chosen as the crossing frequency of the "fast" and "slow" noise components. Then, as a preliminary step, it is convenient to use (5) to characterize the bump in terms of average frequency $f_b$, bandwidth $B_b$, amplitude $b_b$ and in particular power $P_b$. In addition, $P_b$ and Tab. 1 allow estimating the bump contribution to AVAR and to predict the improvement due to filtering. Finally, $A_{F\min}$ is calculated using (9) by comparing $P_b$ with the noise power below $f_h$. The filter should then have $A_F > A_{F\min}$ for $f_b - B_b/2 < f < f_b + B_b/2$.

As an example, we consider the realistic case of an optical link where the phase is sampled at $\tau_0$ = 10 ms, with a bump at $f_b$ = 21.5 Hz whose bandwidth $B_b$ is 22.3 Hz. Let us consider also that $f_h$ is 5 Hz and that, from (10), $A_{F\min}$ is 26.5 dB. Let us compare a moving average filter of duration $1/(2f_h)$ = 100 ms with a more selective filter, such as a sinc filter [33], whose temporal support is truncated at $\pm 5/f_h$, = $\pm 1$ s, 20 times wider than the moving average. Both filters have bandwidth $f_h$ and are normalized so that $|H_F(0)|$ = 1. Using (10), we can calculate their average attenuation $A_F$: −17.9 dB for the moving average and 55.6 dB for the truncated sinc filter. It is thus straightforward to choose the latter, as the moving average is inadequate. We achieve the same conclusion graphically by plotting the transfer functions of the two filters together with the filter specifications (Fig. 4).

In this example, for simplicity, we used a truncated sinc filter. However, a wide variety of digital filter algorithms is available in the most common libraries for digital signal processing, whose parameters can be tailored on the specific experiment.

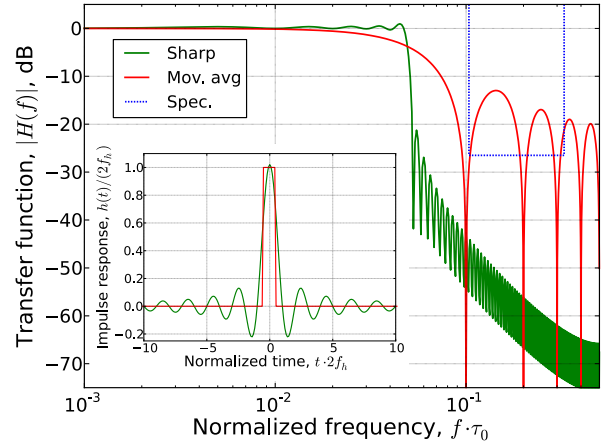

Figure 4: the transfer function plot shows that the selectivity of the truncated sinc filter (in green) is significantly higher than the moving average filter (in red) with the same bandwidth $f_h = 0.05/\tau_0$. This is obtained at the expense of the impulse response duration ($10/f_h$ in the example) as shown by the inset. The selectivity is proportional to the duration of the impulse response and can be changed according to the required attenuation (in blue).

It is advisable to apply filters iteratively with progressively decreasing $f_h$. After each filtering stage, the amount of data can be resized according to the Nyquist theorem; this dramatically reduces the storage and computational requirements. Once the filtering and decimation process is completed, i.e. the optimal $f_h$ has been obtained, the Allan deviation can be computed.

In the example we have shown, $f_h$ was intentionally chosen very close to the bump. This is a worst-case scenario regarding the requirements on filter selectivity. Another significant example is when $f_h \ll f_b$. In this case, even a low selectivity filter as the moving average may guarantee the required attenuation, thanks to the wider transition bandwidth. A deeper insight will be given in Sec. IV, where we compare the numerical results obtained with these truncated sinc and moving average filters on an experimental case.

IV. APPLICATION TO AN EXPERIMENTAL CASE

Let us consider a practical case in which the described analysis can be applied. Initially, we will describe how to evaluate the performance of a coherent optical link and then we will consider a hypothetical clock comparison over this link. For completeness, we will describe both the cases where optical or microwave clocks are compared.

We analyze data of the coherent optical link developed by the National Metrology Institute in Italy (INRIM). This link has a total length of 642 km and connects many national research facilities of Italy. Its metrological characterization has been performed by looping the link using a single fiber, to have both ends in the same laboratory. The total link length is then 1284 km. The experimental realization and the link validation are detailed in [3]. Both the electronics apparatus and the measurement instruments are entirely composed of commercial devices and off-the-shelf components. In particular, we realized fully-analog phase-locked loops with clean-up voltage-controlled oscillators; the data are collected using a dead-time free phase/frequency counter [26] in the phasemeter mode at 1 kHz sampling frequency. The set of data we analyze here lasts 20000 s and contains 20 MSa (about 1 GB of text data).

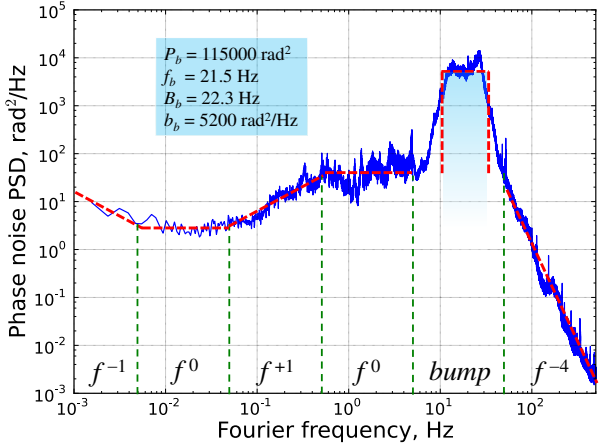

Fig. 5: The phase noise power spectral density (PSD) of the 1284 km optical link. Dashed lines represent the extrapolated noise components. The equivalent bump and its parameter are reported as well.

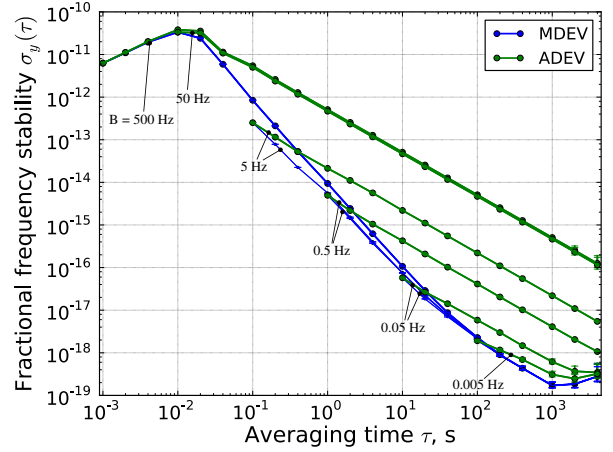

Fig. 6: The ADEV and MDEV (green and blue thin lines respectively) of the 1284 km link with different measurement bandwidths and the unfiltered MDEV (thick blue line). The effect of $f_h$ on MDEV is shown by the thin blue lines.

The first aspect to consider is the power spectral density of the stabilized link phase noise, shown in Fig. 5. It contains many cases of interest: for $f > 50$ Hz, the noise is decreasing quickly ($b_{-4} = 10^8$ rad$^2$Hz$^3$); for 5 Hz $< f <$ 50 Hz, there is a strong bump of 340 rad$_{rms}$; for 0.5 Hz $< f < 5$ Hz the noise is dominated by WPN ($b_0 = 50$ rad$^2$/Hz); for 50 mHz $< f < 0.5$ Hz there is BPN ($b_{+1} = 75$ rad$^2$/Hz$^2$); for 5 mHz $< f < 50$ mHz, the spectrum is flat ($b_0 = 3$ rad$^2$/Hz); at lower frequencies, we find FPN ($b_{-1} = 0.02$ rad$^2$), and then, not shown in figure, the signature of environmental effects, such as temperature variations and transients.

The bump is centered at $f_b = 21.5$ Hz with bandwidth $B_b = 22.3$ Hz and amplitude $b_b = 5200$ rad$^2$/Hz. The power $P_b$ is 115000 rad$^2$ and its contribution ($4.68 \times 10^{-13}$ at 1 s) is by far the main component of the full bandwidth ADEV (the square root of AVAR). The calculated ADEV at 1 s ($f_h = 500$ Hz) agrees at the 0.1% level with the value predicted by Tab. 1.

For the ADEV computation, the most problematic spectral region is between 5 mHz and 50 Hz. To show the effect of each noise type, we apply the proposed method by progressively reducing the bandwidth from 500 Hz to 5 mHz in steps of a factor 10. The results are shown in Fig. 6, together with those obtained with the MDEV. We note that the ADEV is only shown for $\tau \geq 1/(2f_h)$, where it is not affected by the filtering process.

It can be noticed that the bandwidth reduction from 500 Hz to 50 Hz has no impact on the ADEV estimation, as in this spectral region the spectrum decreases quickly. With filtering, the ADEV estimation improves progressively, in good agreement with the prediction expected from the formulae in Table 1. A factor 20 improvement is obtained when the bandwidth is reduced from 50 Hz to 5 Hz, i.e. when the bump is filtered out. Further filtering still improves the ADEV depending on the noise type: a factor 5 ($\alpha = 0$), then a factor 7.2 ($\alpha = +1$) and in the end a factor 2. In the last curve, obtained with $f_h = 5$ mHz, the fast noise has been completely removed, and the ADEV can be considered representative of the link noise. Thus, it is possible to conclude that the link has a stability of $3 \times 10^{-19}$ at 1000 s with $f_h = 5$ mHz.

We note that the filtered ADEV provides better results than the MDEV for $\tau = 1/(2f_h)$, just because the $h_F$ filter is more selective than the moving average embedded in the MDEV. This is particularly evident for $f_h = 5$ Hz, where the bump has been filtered out. At 100 ms, it provides an estimation that is 10 dB lower. We note as well that also the MDEV benefits from the filtering process (thin blue lines). It is interesting to take a closer look to the case of $f_h = 5$ Hz: the unfiltered MDEV takes about two decades (from 100 ms to 10 s) to reach the estimation of the filtered MDEV. In the other cases, the difference is not significant, since the MDEV algorithm well manages noise processes with $\alpha \leq +1$.

## V. CLOCKS COMPARISON AND DISSEMINATION

Once implemented and characterized, the link can be used to transfer stable and accurate frequency references such as those generated by atomic clocks. Currently, our clocks ensemble at INRIM is composed of several hydrogen masers, a Cs cryogenic fountain with an accuracy of $2.5 \times 10^{-16}$ and a stability of $2.0 \times 10^{-13} \tau^{-1/2}$ [34], and an Yb lattice clock under development [35]. The primary Cs standard is disseminated through the optical link via an ultrastable laser, whose frequency is referred to the Cs frequency through an optical comb. Evidently, the same can be done using the optical clock as a reference. The main difference in the two cases relies in the frequency stability performance, being at the level of $2.0 \times 10^{-13} \tau^{-1/2}$ for a microwave clock [34], and as low as $3.2 \times 10^{-16} \tau^{-1/2}$ for an optical clock [35-36].

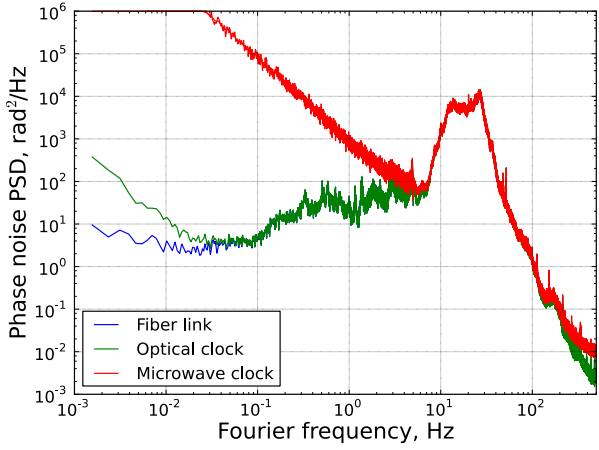

Fig. 7: the phase noise power spectral density of the bare link (blue line) and of a link that disseminates the signals of the simulated microwave and optical clocks in the remote laboratory (red and green line respectively).

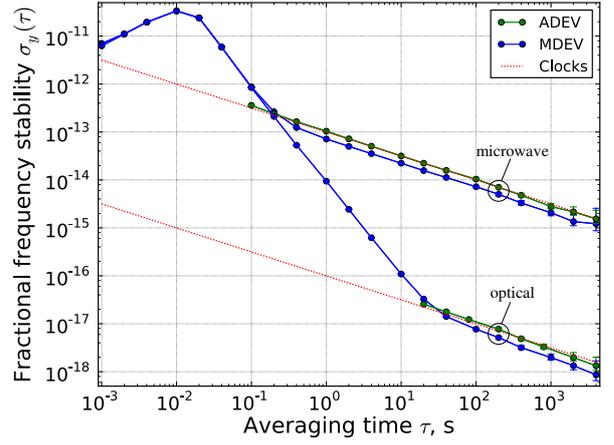

Fig. 8: the stability of the clock signals in the remote laboratory when either the MDEV (blue lines) or the ADEV associated with filtering (green lines) are used. The uppermost curves represent the case of the microwave clock; in this case the ADEV is computed with $f_h$ = 5 Hz. The lower curves are related to the optical clock; in this case the ADEV is computed with $f_h$ = 25 mHz. The dotted red lines represent the stability of the two clocks.

In this section, we consider both the optical and microwave dissemination along an optical link. In particular, each clock is simulated superimposing to the actual link noise a phase random walk obtained by integrating a WFN with Gaussian distribution; for the sake of clarity, here we assume a stability of $1\times10^{-13}\ \tau^{-1/2}$ and $1\times10^{-16}\ \tau^{-1/2}$ for the microwave and optical clocks respectively. Then, the data have been processed by the anti-aliasing filter $h_F$, as explained in the previous sections.

In Fig. 7, we show the phase noise of the signal received by the remote observer. It is evident from the $f^{-2}$ behavior of the spectrum that the information related to the clock arises for frequencies lower than 5 Hz for the microwave clock and 25 mHz for the optical clock. For higher frequencies, the spectrum is dominated by the residual phase noise of the link. In agreement with (3), these frequencies are chosen as $f_h$ in the two cases, so that the link contribution is filtered out and the data contain the clock information only.

Then, we computed the ADEV ($f_h = f_{hopt}$), the MDEV ($f_h$ = 500 Hz) and the theoretical deviations for the two clocks. The results are shown in Fig. 8. In the case of the optical clock, the remote user can recover the clock stability after 20 s of averaging time without bias.

On the other hand, the MDEV provides a −3 dB biased estimate of the clock stability and in addition, it is representative of the clock stability only after 40 s. The advantage in the case of the microwave clock is more evident. Here the filtered ADEV represents the clock stability at the remote site from 100 ms, while the full bandwidth MDEV carries the information of the clock stability from an averaging time of 400 ms. We point out that, at 100 ms, the ADEV well estimates the clock stability, while the MDEV estimation is 7.5 dB higher. This is because, in this case, $f_h$ is close to the bump, which is rejected effectively by the anti-aliasing filter.

As a last consideration, we show in more detail the benefit of sharp filtering on the link noise. The filtering process used to estimate the link performance is the result of cascaded application of a truncated sinc filter with $f_h$ = 0.05 $f_s$. After each filtering stage, data have been decimated by a factor 10. Fig. 9 shows the link phase noise when filtered with the sinc filter and with a moving average. From the picture, it can be seen that the moving average is not selective enough to completely reject the link noise for $f > f_h$, thus affecting the ADEV computation.

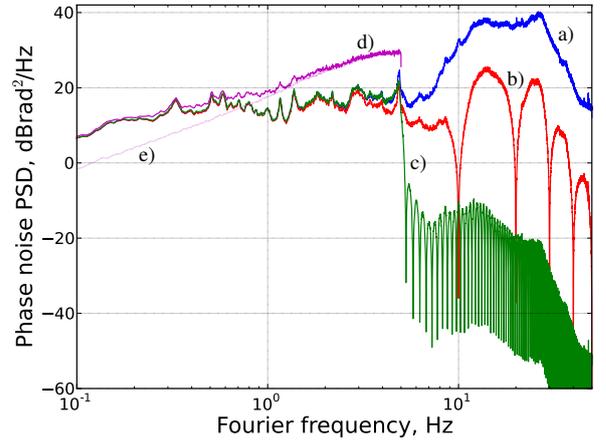

Figure 9: the unfiltered (blue line, a) and filtered link noise when either a moving average (red line, b) or a sinc filter (green line, c) are applied, both with the same $f_h$ = 5 Hz. The thin purple line (d) represents the PSD of data corresponding to the red curve after decimation. The effect of aliasing is clearly visible between 1 Hz and 5 Hz. The dotted line, e) reports the aliasing component and demonstrates its slope.

In addition, care must be taken in the decimation process of simply averaged data, as it could lead to a peculiar type of aliasing. Typically, one expects aliasing to degrade the white noise level; if so, it would completely bury the link noise at low frequency. However, thanks to the first order zeroes of the moving average transfer function (at $f = 2mf_h$, with $m = 1, 2…$), the aliasing contributes as $f^2$ for $f \ll f_h$, therefore becoming negligible with respect to the link noise as shown by Fig. 9. A comparison of the ADEV obtained with sinc filtering and moving average is shown in Fig. 10. A higher stability is noticed when the moving average is used: this is expected, given the lower selectivity of this filter, nevertheless, at 5 mHz, far from the bump, the results obtained are very close to the ones calculated by using the sharp filter.

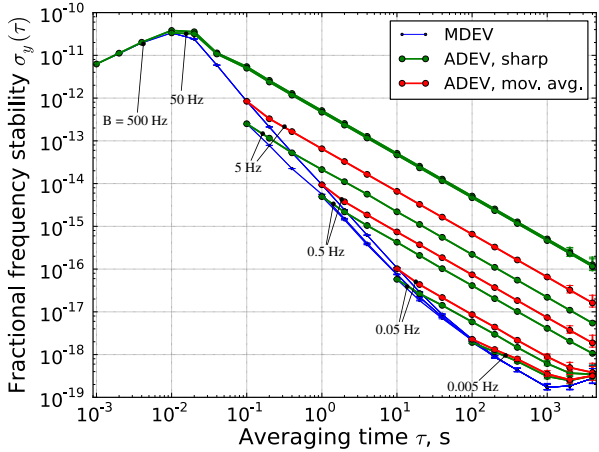

Figure 10: the ADEV computed from sinc-filtered data (green curves) or moving-average-filtered data (red curves) for different bandwidths.

## VI. CONCLUSION

In this paper, we have shown how to deal with aliasing in the computation of the Allan deviation. Our analysis is general and can be applied in many contexts; in this work, we focus on the characterization of optical frequency links, which significantly suffer for this problem. In order to treat properly the problem, we have modeled the link phase noise and analytically derived the ADEV as a function of the measurement bandwidth $f_h$, showing how this parameter can be regarded as an additional degree of freedom. We calculated the optimum measurement bandwidth and the minimum attenuation required by the anti-aliasing filter to reject high-frequency noise components and detailed how to determine these parameters.

Aliasing is a serious issue in the optical phase measurement, especially when frequency counters are used to this purpose. We gave an overview of the possible limitations and on how to cope with them.

The proposed method allows the use of the AVAR as a statistical estimator in fiber-based remote clocks comparisons as well as for bare evaluation of the ultimate optical links performances. This approach preserves the continuity with the existing literature, where the AVAR is traditionally used in clock uncertainty evaluation.

We applied this technique to analyze experimental data of the 1284 km optical fiber link for frequency dissemination developed by the National Institute of Metrology in Italy. We evaluated the link performance and we simulated two clock comparisons over this fiber. Overall, we have shown how the ADEV can be effectively used to characterize the link stability.

This approach can be useful in view of an increasing number of clock comparisons via optical fiber, also considering that it can be readily implemented in commercial devices such as dead-time free phase/frequency counters.

ACKNOWLEDGMENTS

The authors thank the Optical Link team at INRIM for useful discussion and for allowing use of the experimental data, Marco Pizzocaro for useful suggestions and careful reading of the Manuscript and Carolina Cárdenas for her help in data computation.

This work was supported by the Italian Ministry of Research and Education under the LIFT project of the Progetti Premiali program and by the European Metrology Research Program (EMRP project SIB55-ITOC). The EMRP is jointly funded by the EMRP participating countries within EURAMET and the European Union.